\documentclass[12pt]{iopart}
\usepackage{hyperref}
\usepackage{iopams}

\begin{document}
\title[Double layer]{Double layer from least action principle}
\author{V A Berezin, V I Dokuchaev, Yu N Eroshenko \\ and A L Smirnov}
\address{Institute for Nuclear Research of the Russian Academy of Sciences, \\
60th October Anniversary Prospect 7a, 117312 Moscow, Russia}
\eads{\mailto{berezin@inr.ac.ru}, \mailto{dokuchaev@inr.ac.ru}, \mailto{eroshenko@inr.ac.ru}, \mailto{smirnov@inr.ac.ru}}

\begin{abstract}
We derived the equations for the double layers in Quadratic Gravity, using solely the least action principle. The advantage of our approach is that, in the process of calculation, the $\delta'$-function does not appear at all, and the $\delta$-functions appear for a moment and are mutually canceled prior to integration. We revealed the peculiar structure of the obtained equations, namely, that the surface energy-momentum tensor of the matter fields (constituents of the thin shells) does not play a role in the determination of the trajectory of the double layer. Also, we suggested that the space-like double layers may provide us with the adequate description of the creation of the universe from the black hole singularity. The related topics, including the Gauss-Bonnet term and $F(R)$-theories, are shortly discussed. 
\end{abstract}

\section{Introduction}

The role of the exact solutions in the understanding of physical phenomena can hardly be overestimated. Since the field equations of any relativistic theory of gravitation is highly nonlinear, the search for the solutions becomes the very uneasy task. That's why the investigation of singular distributions of matter fields plays an important role. In the case of General Relativity the singular hypersurfaces were first investigated by W.~Israel \cite{Israel66,Israel67,Cruz67}. He considered the $\delta$-function-like matter energy-momentum tensor and found that the extrinsic curvature tensor, describing embedding of the singular hypersurface into the 4-dimensional space-time, undergoes a jump. The curvature in such a situation also exhibits the $\delta$-function behavior. Note that a jump in curvature can be interpreted as the gravitational shock wave. In Quadratic Gravity the situation is more subtle. If one would allow the $\delta$-function behavior of the curvature, then, in generic case, there would appear the $\delta^2$-terms already in the Lagrangian, what is forbidden in the conventional theory of distributions. Therefore, the curvature may have, at most, jumps at the singular hypersurface. Then, since the field equations in Quadratic Gravity are of second order in derivatives of the curvature (i.e.,  fourth order in derivatives of the metric tensor), the appearance of the terms proportional both to $\delta$-function and its derivative, $\delta'$-function, is inevitable. This, by definition, is what we call the ``double layer''.

Thus, we see, that in Quadratic Gravity the gravitational shock waves are dressed in the double layer uniform. The thin shells in the quadratic gravity were investigated by H. -H. von Borzeszkowski and V. P. Frolov \cite{Frolov80}. J. M. M. Senovilla recognized that the double layer appear as well and derived the double layer equations in the bulk (see details in \cite{Senovilla13,Senovilla14,Senovilla15,Senovilla16,Senovilla17,Senovilla18}). He did this in a quite general form. In the present paper we would like to derive the double layer equations from the least action principle only and will use as much as possible the Gauss normal coordinate system associated with the singular hypersurface, in order to make the results more physically readable and ready for the applications.

\section{Preliminaries}

We start from definitions. The action integral for Quadratic Gravity is
\begin{equation}
S_{\rm gr}=\int\limits_{\Omega}\mathcal{L}_2\sqrt{-g}\,d^4x
\end{equation}
with the Lagrangian
\begin{equation}
\mathcal{L}_2=\alpha_1R_{\mu\nu\lambda\sigma}R^{\mu\nu\lambda\sigma}+\alpha_2R_{\mu\nu}R^{\mu\nu}+\alpha_3R^2+\alpha_4R+\alpha_5\Lambda,
\end{equation}
where $R^{\mu}_{\phantom{\mu}\nu\lambda\sigma}$ is Riemann curvature tensor
\begin{equation}
R^{\mu}_{\phantom{\mu}\nu\lambda\sigma}=\frac{\partial \Gamma^\mu_{\nu\sigma}}{\partial x^\lambda}-\frac{\partial \Gamma^\mu_{\nu\lambda}}{\partial x^\sigma}+\Gamma^\mu_{\varkappa\lambda}\Gamma^\varkappa_{\nu\sigma}-\Gamma^\mu_{\varkappa\sigma}\Gamma^\varkappa_{\nu\lambda},
\end{equation}
$R_{\mu\nu}$ is Ricci tensor
\begin{equation}
R_{\mu\nu}=R^{\lambda}_{\phantom{\mu}\mu\lambda\nu},
\end{equation}
and $R$ is the curvature scalar
\begin{equation}
R=R^\lambda_{\lambda}.
\end{equation}

We are working in the framework of Riemannian geometry, so, the dynamical variables are components of the metric tensor $g_{\mu\nu}$,
\begin{equation}
ds^2=g_{\mu\nu}dx^\mu dx^\nu,
\end{equation}
the indices are raised and lowered by $g_{\mu\nu}$ and its inverse $g^{\mu\nu}$ ($g^{\mu\nu} g_{\nu\lambda}=\delta^\mu_\lambda$), and
the connections are just the Christoffel symbols
\begin{equation}
\Gamma^\lambda_{\mu\nu}=\frac{1}{2}g^{\lambda\varkappa}\left(g_{\varkappa\mu,\nu}+g_{\varkappa\nu,\mu}-g_{\mu\nu,\varkappa}\right),
\end{equation}
where comma (,) denotes a partial derivative.

The total action is given by the sum of the gravitational action, $S_{\rm gr}$, and the action for the matter fields, $S_{\rm m}$,
\begin{equation}
S_{\rm tot}=S_{\rm gr}+S_{\rm m},
\end{equation}
and the least action principle requires the vanishing of the variation of the total action inside the volume of integration, provided the values of the dynamical variables on the overall boundary, $\partial\Omega$, are fixed, i.e. $\delta g_{\mu\nu}|_{\partial\Omega}=0$. Thus,
\begin{equation}
\delta S_{\rm tot}=0.
\end{equation}
The variation of $S_{\rm m}$ gives us, by definition, the energy-momentum tensor, $T_{\mu\nu}$, of the matter field.
\begin{eqnarray}
\delta S_{\rm m}&\equiv&\frac{1}{2}\int 
T_{\mu\nu}(\delta g^{\mu\nu})\sqrt{-g}\,d^4x =
-\frac{1}{2}\int T^{\mu\nu}(\delta g_{\mu\nu})\sqrt{-g}\,d^4x.
\end{eqnarray}

We are interested here in the situation when the energy-momentum tensor has either different behavior on two different sides of some hypersurface $\Sigma_0$, or one has different vacua separated by such a hypersurface, with the gravitating source concentrated on it. Accordingly, we assume that the energy-momentum tensor has the form
\begin{equation}
T^{\mu\nu}=S^{\mu\nu}\delta(n)+T^{\mu\nu}(+)\theta(n)+T^{\mu\nu}(-)\theta(-n),
\end{equation}
where $\delta(n)$ is the Dirac's $\delta$-function, $\theta(n)$ is the Heaviside step function, and $n=0$ is the equation of the hypersurface $\Sigma_0$, $n(x^\mu)=0$, such an equation being, of course, different in $(+)$ (``outer'') and ($-$) (inner) regions of the bulk. The tensor $S^{\mu\nu}$ is called the surface energy-momentum tensor (we assume that no derivatives of $\delta$-function should appear, i.e., no double wall with opposite signs of energy, no exotic matter on $\Sigma_0$). The matching conditions for the solutions in $\pm$-regions are the equations for determining the surface $\Sigma_0$ itself.

The hypersurface $\Sigma_0$ is characterized by 3-dimensional metric tensor $\gamma_{ij}$ (in what follows we adopt the 4-dimensional signature $(+---)$, the Greek indices take values $\{0,1,2,3\}$, while the Latin ones --- $\{0,2,3\}$ or $\{1,2,3\}$) and the extrinsic curvature tensor $K_{ij}$, which describes the embedding of a 3-dimensional surface into 4-dimensional space-time. It is well known that, by suitable coordinate transformations (separately in ($+$)- and ($-$)-regions) the metric tensor can be made continuous on any hypersurface. We will consider here the non-null hypersurface $\Sigma_0$. In such a case the famous example is the Gauss normal coordinate system associated with $\Sigma_0$. It reads
\begin{equation}
ds^2=\epsilon dn^2+\gamma_{ij}dx^idx^j,
\end{equation}
\begin{equation}
\epsilon=\pm1, \qquad \gamma_{ij}=\gamma_{ij}(u,x), 
\end{equation}
where $\epsilon=+1$, if $\Sigma_0$ is space-like and $\epsilon=-1$ if it is time-like, and $n$ is a coordinate along the outward normal to $\Sigma_0$ (i.e., it goes from ($-$)- to ($+$)-region). We will use this coordinates as the basic ones just nearby the hypersurface $\Sigma_0$. The extrinsic curvature tensor in the Gauss normal coordinate equals
\begin{equation}
K_{ij}=-\frac{1}{2}\gamma_{ij,n}.
\end{equation}
Having continuous metric tensor on $\Sigma_0$, what can be said about the extrinsic curvature tensor $K_{ij}$? The answer is different for General Relativity and Quadratic Gravity.

In General Relativity the gravitational field equations, the Einstein equations are of the second order in derivatives of the metric tensor. Hence, the appearance of $\delta$-function in the matter distribution leads to its appearance in the second derivatives of the metric tensor, i.e., in the Riemann tensor. Therefore, their first derivatives, i.\,e., connection coefficients should have a jump across the matching hypersurface $\Sigma_0$,
\begin{equation}
[g_{\mu\nu,\lambda}]|_{\Sigma_0}\neq 0,
\end{equation}
where $[~]=(+)-(-)$. Thus, 
\begin{equation}
[K_{ij}]\ne0,
\end{equation}
in such a case $\Sigma_0$ is called singular hypersurface, or the thin shell. The corresponding matching conditions connecting these jumps with the surface energy-momentum tensor $S_{ij}$ were first derived by W.~Israel \cite{Israel66,Israel67,Cruz67}. Note, if there is no $\delta$-function in the matter distribution, but only a jump, then the corresponding jump in the curvature describes the gravitational shock wave accompanied by the shock wave in the matter.

In Quadratic Gravity the jumps in the connections and, consequently, the $\delta$-functions in the curvature, would result in the $\delta^2$-terms in the Lagrangian, what is forbidden in the conventional theory of distributions. To avoid this, one has to impose the Lichnerowicz conditions (see details and references in \cite{Lake}), namely, the first derivatives of the metric tensor must be continuous at the singular hypersurface, 
\begin{equation}
[g_{\mu\nu,\lambda}]=0,
\end{equation}
hence
\begin{equation}
[K_{ij}]=0.
\end{equation}
The field equations now are of the fourth order in derivatives of the metric tensor (of the second order in derivatives of the curvature). Therefore, we have two possibilities. Either, the curvature is continuous at $\Sigma_0$, then its second derivative may contain, at most, the $\delta$-function, which would become a counterpart of the $\delta$-function term in the energy-momentum tensor. This is the thin shell situation, but the equations will be quite different from the Israel's ones. Or the curvature undergoes a jump at $\Sigma_0$, then, its first derivative will contain the $\delta$-function term, while the second derivative --- $\delta'$ (the derivative of the $\delta$-function). In this case the singular hypersurface is called the double layer. The equations of the double layer in Quadratic Gravity were first derived by J.~M.~M.~Senovilla.

The aim of the present work is to derive the equations for the double layer straight from the least action principle.

Note that now the jump in the curvature, describing the gravitational shock wave, may be or may not be accompanied by the shock wave in the matter distribution. Thus, in Quadratic Gravity the pure gravitational shock wave may exist.

\section{Variation process}

We are interested here in the field equation on the singular hypersurface $\Sigma_0$, which play the role of the matching conditions for the solutions in the ($\pm$)-regions (in the bulk) where the matter energy-momentum tensors have different structure, or with different vacuum solutions there. Therefore, we assume that the field equations in the bulk are already fulfilled,
\begin{equation}
\delta S_{\rm tot}(\pm)=0.
\end{equation}
So,
\begin{equation}
\delta S_{\rm gr}(\Sigma_0)=-\delta S_{\rm m}(\Sigma_0)=\frac{1}{2}\int\limits_{\Sigma_0} T_{\mu\nu}(\delta g^{\mu\nu})\sqrt{|\gamma|}\,d^3x,
\end{equation}
where $\gamma$ is the determinant of the metric $\gamma_{ij}$ on singular hypersurface $\Sigma_0$, introduced above. Due to the Lichnerowicz conditions, we may have only jumps at $\Sigma_0$ in the curvature, no the Dirac's $\delta$-functions. Therefore, there will not be direct contributions to the surface integral from the Lagrangian $\mathcal{L}_2$. So, we have to consider the variation of the full gravitational action, but leaving only those terms that contribute to the surface integral. For this very reason we will omit all the terms that proportional to $\delta g_{\mu\nu}(g^{\mu\nu})$. Following this rule we obtain
\begin{eqnarray}
\delta S_{\rm gr}&=&\!\!\!\int\!\left(
(\delta \mathcal{L}_m)+\frac{1}{2}\mathcal{L}_m g^{\mu\nu}(\delta g_{\mu\nu})
\right)\sqrt{-g}\,d^4x\:\to \!\int
(\delta \mathcal{L}_m) \sqrt{-g}\,d^4x, \\
\delta S_{\rm gr}&\to& \!\!\int\!\Bigl(
2\alpha_1 R_{\mu}^{\phantom{\mu}\nu\lambda\sigma}(\delta R^{\mu}_{\phantom{\mu}\nu\lambda\sigma})+2\alpha_2R^{\mu\nu}(\delta R_{\mu\nu}) 
\nonumber \\
&+& \!2\alpha_3 Rg^{\mu\nu}(\delta R_{\mu\nu})+
\alpha_4 g^{\mu\nu}(\delta R_{\mu\nu}) \Bigr)\sqrt{-g}\,d^4x.
\end{eqnarray}
The choice of just these combinations of the variations is dictated by the existence of nice formulas, first found by Palatini \cite{Palatini}
\begin{equation}
\delta R^{\mu}_{\phantom{\mu}\nu\lambda\sigma}=(\delta\Gamma^\mu_{\nu\sigma})_{;\lambda}-(\delta\Gamma^\mu_{\nu\lambda})_{;\sigma},
\end{equation}
where a semicolon $(;)$ denotes a covariant derivative with respect to the metric connections $\Gamma$ (Christoffel symbols) (note, that $\delta\Gamma^\mu_{\nu\lambda}$ is a tensor)\footnote{V.A.B. is indebted to Prof. Friedrich Hehl for indication the author of this remarkable relation.}. It is easy to show that the term with $\alpha_4$ (Hilbert action) does not contribute to the integral over $\Sigma_0$. Indeed, using the Palatini formula, we obtain
\begin{eqnarray}
\alpha_4\int\! g^{\mu\nu}(\delta R_{\mu\nu})\sqrt{-g}\,d^4
x=\alpha_4\int\! \left\{g^{\mu\nu}
\Bigl((\delta\Gamma^\lambda_{\mu\nu})_{;\lambda}
-(\delta\Gamma^\lambda_{\mu\lambda})_{;\nu}\Bigr)\right\}
\sqrt{-g}\,d^4x
\nonumber \\
=\alpha_4\int\! \Bigl\{\Bigl((g^{\mu\nu}(\delta\Gamma^\lambda_{\mu\nu})_{;\lambda}-(g^{\mu\nu}\delta\Gamma^\lambda_{\mu\lambda})_{;\nu}\Bigr)\Bigr\}\sqrt{-g}\,d^4x.
\end{eqnarray}
Since, for any vector $l^\sigma$ one has $l^\sigma_{;\sigma}\sqrt{-g}=(\sqrt{-g}l^\sigma)_{;\sigma}$, then the above expression is just the linear combination of the full derivatives and, by Stokes' theorem, equals 
\begin{equation}
-\alpha_4\int\! \Bigl(g^{\mu\nu}[\delta\Gamma^\lambda_{\mu\nu})] - g^{\mu\lambda}[\delta\Gamma^\nu_{\mu\nu})]\Bigr)dS_\nu,
\end{equation}
where the vector $dS_\nu$ is directed along the normal coordinate $n$, (from the ($-$)-region to the ($+$)-region). The change of the sign in front of the integral occurs due to our definition $[~]=(+)-(-)$.

The Lichnerowicz conditions require $[\delta\Gamma^\lambda_{\mu\nu}]=0$, that is why there is no contribution from the $\alpha_4$-term into the surface integral. So, we are left with
\begin{eqnarray}
\delta S_{\rm gr}&\to& 2\int\Bigl\{
\alpha_1 R_{\mu}^{\phantom{\mu}\nu\lambda\sigma}(\delta R^{\mu}_{\phantom{\mu}\nu\lambda\sigma})
+\alpha_2R^{\mu\nu}(\delta R_{\mu\nu})
\nonumber \\
&+& \!\! \alpha_3 Rg^{\mu\nu}(\delta R_{\mu\nu})
\Bigr\}\sqrt{-g}\,d^4x.
\end{eqnarray}
It is convenient to consider $\alpha_1$, $\alpha_2$, and $\alpha_3$ parts separately.

\subsection{$\alpha_1$ --- patch}
\label{alpha1patch}

Let us denote the $\alpha_1$-patch by $\delta S_{\rm gr}(\alpha_1)$. Substituting the Palatini formula gives
\begin{eqnarray}
\delta S_{\rm gr}(\alpha_1)&=&2\!\alpha_1\int \! R_{\mu}^{\phantom{\mu}\nu\lambda\sigma}(\delta R^{\mu}_{\phantom{\mu}\nu\lambda\sigma})\sqrt{-g}\,d^4x
\nonumber \\
&=&2\alpha_1\!\int\! R_{\mu}^{\phantom{\mu}\nu\lambda\sigma}
\Bigl(
(\delta\Gamma^\mu_{\nu\sigma})_{;\lambda}-(\delta\Gamma^\mu_{\nu\lambda})_{;\sigma}
\Bigr)\sqrt{-g}\,d^4x
\nonumber \\
&=&4\alpha_1\!\int\! R_{\mu}^{\phantom{\mu}\nu\lambda\sigma}(\delta\Gamma^\mu_{\nu\sigma})_{;\lambda}\sqrt{-g}\,d^4x.
\end{eqnarray}   
The transition to the last line reflect a skew symmetric property of the Riemann curvature tensor. The next step is extracting the full derivative,
\begin{equation}
\delta S_{\rm gr}(\alpha_1)=4\alpha_1\!\int \Bigl\{
\left( R_{\mu}^{\phantom{\mu}\nu\lambda\sigma}(\delta\Gamma^\mu_{\nu\sigma})\right)_{;\lambda}-
R_{\mu\phantom{\mu\mu\mu};\lambda}^{\phantom{\mu}\nu\lambda\sigma}(\delta\Gamma^\mu_{\nu\sigma})
\Bigr\}\sqrt{-g}\,d^4x.
\end{equation}
Here, for the first time, the $\delta$-function shows itself. Indeed, since
\begin{equation}
R_{\mu}^{\phantom{\mu}\nu\lambda\sigma}=R_{\mu}^{\phantom{\mu}\nu\lambda\sigma}(+)\theta(n)+R_{\mu}^{\phantom{\mu}\nu\lambda\sigma}(-)\theta(-n)
\end{equation}
and $[\delta\Gamma^\mu_{\nu\sigma}]=0$, one has
\begin{equation}
\left( R_{\mu}^{\phantom{\mu}\nu\lambda\sigma}(\delta\Gamma^\mu_{\nu\sigma})\right)_{;\lambda}=
[R_{\mu}^{\phantom{\mu}\nu\lambda\sigma}](\delta\Gamma^\mu_{\nu\sigma})\delta(n)n_{,\lambda}+\ldots
\end{equation}
($n(x^\mu)=0$ is an equation for $\Sigma_0$). We see, however, that exactly the same expression appears in the second term in the integrand, but with the opposite sign! And such a situation will be repeated once and once more in the subsequent calculations. The origin of this, of course, lies in the absence of the $\delta$-functions in the Quadratic Gravity Lagrangian $\mathcal{L}_2$. So, we are left with the integrals over $(\pm)$-regions only, and can safely implement the Stokes' theorem,
\begin{eqnarray}
\delta S_{\rm gr}(\alpha_1)=&&-4\alpha_1\int\limits_{\Sigma_0}
[R_{\mu}^{\phantom{\mu}\nu\lambda\sigma}](\delta\Gamma^\mu_{\nu\sigma})
\sqrt{-g}dS_\lambda
\nonumber \\
&&-4\alpha_1\int\limits_{(\pm)}
R_{\mu\phantom{\mu\mu\mu};\lambda}^{\phantom{\mu}\nu\lambda\sigma} (\delta\Gamma^\mu_{\nu\sigma})
\sqrt{-g}\,d^4x.
\end{eqnarray} 
(Note, again, the change of the sign in front of the surface integral.)

Consider, first, the surface integral. In the Gauss normal coordinate system it becomes
\begin{equation}
-4\alpha_1\int\limits_{\Sigma_0}
[R_{\mu}^{\phantom{\mu}\nu n\sigma}](\delta\Gamma^\mu_{\nu\sigma})
\sqrt{|\gamma|}\,d^3x,
\end{equation}
$n$ being the coordinate along the normal to $\Sigma_0$, directing from $(-)$ to $(+)$. Below we enlisted the nonzero components for the Christoffel symbols and jumps in the curvatures in the Gauss normal coordinates (also for future use).
\begin{eqnarray}
\Gamma^n_{ij}=\epsilon K_{ij}, \qquad \Gamma^i_{nj}=-K^i_j,  \qquad  \Gamma^l_{ij}=\tilde\Gamma^l_{ij},
\nonumber \\
\left[R_{ninj}\right]=[K_{ij,n}], \qquad [R_{nn}]=\gamma^{lp}[K_{lp,n}], 
\nonumber \\
\left[R_{ij}\right]=\epsilon[K_{ij,n}], \qquad [R]=2\epsilon K^{lp}[K_{lp,n}].  
\end{eqnarray} 
But, one must be careful with $\delta\Gamma^\lambda_{\mu\nu}$. For the variation of the Christoffel symbol the following expression is valid 
\begin{equation}
\delta\Gamma^\lambda_{\mu\nu}=\frac{1}{2}g^{\lambda\varkappa}\left(
(\delta g_{\varkappa\mu})_{;\nu}+(\delta g_{\varkappa\nu})_{;\mu}-(\delta g_{\mu\nu})_{;\varkappa}
\right).
\end{equation}
Thus, the integrand above becomes
\begin{eqnarray}
[R_{\mu}^{\phantom{\mu}\nu n\sigma}](\delta\Gamma^\mu_{\nu\sigma})&=&
\frac{1}{2}[R^{\mu\nu n\sigma}]\Bigl((\delta g_{\mu\nu})_{;\sigma}
+(\delta g_{\mu\sigma})_{;\nu}-(\delta g_{\nu\sigma})_{;\mu}\Bigr)
\nonumber \\
&=& [R^{\mu\nu n\sigma}](\delta g_{\mu\sigma})_{;\nu}
=\left[R^{nlnp}\right]\Bigl((\delta g_{np})_{;l}-(\delta g_{lp})_{;n}\Bigr),
\end{eqnarray} 
where the known symmetries of the curvature tensor have been extensively used. Since
\begin{eqnarray}
(\delta g_{np})_{;l}=(\delta g_{np})_{|l}+K^r_l(\delta g_{rp})-\epsilon K_{lp}(\delta g_{nn})
\nonumber
\\
(\delta g_{lp})_{;n}=(\delta g_{lp})_{,n}+K^r_l(\delta g_{rp})+K^r_p(\delta g_{lr}),
\end{eqnarray} 
(with vertical line $(|)$ denotes a 3-dimensional covariant derivative with respect to the metric $\gamma_{ij}$) then
\begin{eqnarray}
[R_{\mu}^{\phantom{\mu}\nu n\sigma}](\delta\Gamma^\mu_{\nu\sigma})&=&
g^{il}g^{jp}[K_{lp,n}]\left((\delta g_{nj})_{|i}-(\delta g_{ij})_{,n} \right.
\nonumber
\\
&&\!\!-\epsilon K_{ij}(\delta g_{nn})-K^r_j(\delta g_{ir})\left. \right).
\end{eqnarray}

The problem is that, while in the Gauss coordinate system $g_{nn}=\epsilon=const$ and $g_{ni}=0$, their variations, $\delta g_{nn}$ and $\delta g_{ni}$, are not zero. So, we have to use more general coordinate system in order to deal with them. It seems that the most suitable one in such a case is the following
\begin{equation}
ds^2=g_{nn}dn^2+2g_{ni}dndx^i+g_{ij}dx^idx^j,
\end{equation}
where $n$ is still the coordinate along the normal to the singular hypersurface $\Sigma_0$. It is well known that the metric tensor $\gamma_{ij}$ on any hypersurface $n=const$ equals
\begin{equation}
\gamma_{ij}=g_{ij}-\frac{g_{ni}g_{nj}}{g_{00}}.
\end{equation}
And, of course, after making the variations, one can, again, use the Gauss normal coordinate system. Assuming this is the case, we obtain immediately
\begin{eqnarray}
\delta g_{ij}|_{\Sigma_0}=\delta \gamma_{ij},
\nonumber
\\
\delta (g_{ij,n})|_{\Sigma_0}=\delta (\gamma_{ij,n})=-2\delta K_{ij}.
\end{eqnarray}
Inserting all this stuff into the integrand we obtain for the surface integral in question
\begin{eqnarray}
4\alpha_1\!\int\limits_{\Sigma_0}
[K_{lp,n}]\left(-2g^{il}g^{jp}(\delta K_{ij}) \right.
\nonumber
\\
\left.
+\epsilon K^{lp}\delta (g_{nn})-g^{il}g^{jp}(\delta g_{in})_{|j}+g^{il}K^{jp}(\delta g_{ij})
\right)\sqrt{|\gamma|}\,d^3x.
\end{eqnarray}
The only thing left, is to get rid of the 3-dimensional derivative $(\delta g_{in})_{|j}$. This is an easy exercise because of ``the boundary of the boundary is zero''. The result is
\begin{eqnarray}
4\alpha_1\int\limits_{\Sigma_0}\!\Bigl(
-2g^{il}g^{jp}[K_{lp,n}](\delta K_{ij})+\epsilon K^{lp}[K_{lp,n}](\delta g_{nn})
\nonumber \\
+g^{il}g^{jp}[K_{lp,n|j}](\delta g_{in})
+g^{il}K^{jp}[K_{lp,n}](\delta \gamma_{ij})
\Bigr)\sqrt{|\gamma|}\,d^3x.
\end{eqnarray}

Let us now turn to the remaining volume integral,
\begin{equation}
-4\alpha_1\!\int\!
R_{\mu\phantom{\mu\mu\mu};\lambda}^{\phantom{\mu}\nu\lambda\sigma}
(\delta\Gamma^\mu_{\nu\sigma})\sqrt{-g}\,d^4x.
\end{equation}
Substituting the expression for $\delta\Gamma^\mu_{\nu\sigma}$, we get
\begin{eqnarray}
&&-2\alpha_1\!\int\!
R^{\mu\nu\lambda\sigma}_{\phantom{\mu\mu\mu\mu};\lambda}\left(
(\delta g_{\mu\nu})_{;\sigma}+(\delta g_{\mu\sigma})_{;\nu}-(\delta g_{\nu\sigma})_{;\mu} \right) \sqrt{-g}\,d^4x
\nonumber \\
&=&-4\alpha_1\!\int\!
R^{\mu\nu\lambda\sigma}_{\phantom{\mu\mu\mu\mu};\lambda}
(\delta g_{\mu\sigma})_{;\nu}
\sqrt{-g}\,d^4x \to
\nonumber \\
&&\to
4\alpha_1\int\limits_{\Sigma_0}
[R^{\mu n\lambda\sigma}_{\phantom{\mu\mu\mu\mu};\lambda}](\delta g_{\mu\sigma})
\sqrt{|\gamma|}\,d^3x 
\nonumber
\\
&=&4\alpha_1\int\limits_{\Sigma_0}
[R^{ln\lambda\sigma}_{\phantom{\mu\mu\mu\mu};\lambda}](\delta g_{l\sigma})
\sqrt{|\gamma|}\,d^3x
\end{eqnarray}
(it is easy to show that $R^{nn\lambda\sigma}_{\phantom{\mu\mu\mu\mu};\lambda}=0$).

Going further, one finds
\begin{eqnarray}
[R^{ln\lambda\sigma}_{\phantom{\mu\mu\mu\mu};\lambda}](\delta g_{l\sigma})=[R^{ninj}_{\phantom{\mu\mu\mu\mu}|j}](\delta g_{in})
\nonumber \\
+\Bigl(-[R^{ninj}_{\phantom{\mu\mu\mu\mu},n}]
+K^i_l[R^{nlnj}]+K[R^{ninj}] \Bigr) (\delta g_{ij}).
\end{eqnarray}
And
\begin{equation}
[R^{ninj}_{\phantom{\mu\mu\mu\mu},n}]=g^{il}g^{jp}[K_{lp,nn}]+6K^{jp}g^{il}[K_{lp,n}].
\end{equation}
Note, that $[R^{ninj}_{\phantom{\mu\mu\mu\mu},n}]\neq[R^{ninj}_{\phantom{\mu\mu\mu\mu}}]_{,n}$. 

Putting everything together, we obtain the following final result for $\delta S_{\rm gr}(\alpha_1)$
\begin{eqnarray}
\delta S_{\rm gr}(\alpha_1)&=& 4\alpha_1\!\int\limits_{\Sigma_0}\!
\Bigl\{-2g^{il}g^{jp}[K_{lp,n}](\delta K_{ij})
+\epsilon K^{lp}[K_{lp,n}](\delta g_{nn})
\nonumber \\
&&\!+2g^{il}g^{jp}[K_{lp,n|j}](\delta g_{in})
+\!\Bigl(\!-g^{il}g^{jp}[K_{lp,nn}]-4g^{il}K^{jp}[K_{lp,n}]
\nonumber \\
&&\!+\!
Kg^{il}g^{jp}[K_{lp,n}]\Bigr) (\delta \gamma_{ij}) \Bigr\}\sqrt{|\gamma|}\,d^3x.
\end{eqnarray}

\subsection{$\alpha_2$ --- patch}
\label{alpha2patch}

We have already described the subtle points in the previous Subsection~\ref{alpha1patch}. So, here we will present only the main steps  in calculations  and results.

As before, we first transform the  $\alpha_2$ --- patch in the following way 
\begin{eqnarray}
\label{alpha2}
&&2\alpha_2\!\int\!\! R^{\mu\nu}(\delta R_{\mu\nu})\sqrt{-g}\,d^4x \\
&=&2\alpha_2\!\int\!\! R^{\mu\nu}\left((\delta\Gamma_{\mu\nu}^\lambda)_{;\lambda}
-(\delta\Gamma_{\mu\lambda}^\lambda)_{;\nu}\right)\!\sqrt{-g}\,d^4x
 \nonumber \\
&=&2\alpha_2\!\int\! 
\Bigl\{\left(R^{\mu\nu}(\delta\Gamma_{\mu\nu}^\lambda)\right)_{;\lambda}\!
-\left(R^{\mu\nu}(\delta\Gamma_{\mu\lambda}^\lambda)\right)_{;\nu}
\Bigr\} \sqrt{-g}\,d^4x \nonumber \\
&&-\Bigl\{\!
R^{\mu\nu}_{\phantom{\mu\nu}\!;\lambda}(\delta\Gamma_{\mu\nu}^\lambda)
-R^{\mu\nu}_{;\nu}(\delta\Gamma_{\mu\lambda}^\lambda)
\Bigr\} \sqrt{-g}\,d^4x \nonumber \\
&=&-2\alpha_2\!\int\! 
\Bigl\{[R^{\mu\nu}]
(\delta\Gamma_{\mu\nu}^n) - [R^{\mu n}](\delta\Gamma_{\mu\lambda}^\lambda)
\Bigr\}\sqrt{|\gamma|}\,d^3x 
\nonumber \\
&&-2\alpha_2\!\int\limits_{(\pm)}\! 
\Bigl\{\!
R^{\mu\nu}_{\phantom{ij};\lambda}(\delta\Gamma_{\mu\nu}^\lambda)
-R^{\mu\nu}_{\phantom{ij};\nu}(\delta\Gamma_{\mu\lambda}^\lambda)
\Bigr\}\sqrt{-g}\,d^4x. 
\end{eqnarray}
The final result for the surface integral reads
\begin{eqnarray}
\label{alpha2final}
&&2\alpha_2\!\int\limits_{\Sigma_0}\!
\Bigl\{-(g^{il}g^{jp}+g^{ij}g^{lp})[K_{lp,n}](\delta K_{ij})
 \nonumber \\ 
&&+\epsilon K^{lp}[K_{lp,n}](\delta g_{nn}) +
g^{il}g^{jp}[K_{lp,n|j}](\delta g_{in})  \nonumber \\ 
&&+ g^{lp}K^{ij}[K_{lp,n}](\delta \gamma_{ij}) 
\Bigr\} \sqrt{|\gamma|}\,d^3x. 
\end{eqnarray}
The transformation of the remaining volume integral into the integral over the singular hypersurface  $\Sigma_0$ gives us
\begin{eqnarray}
\label{alpha2volume}
&&-2\alpha_2\!\int\!
\Bigl\{  
R^{\mu\nu}_{\phantom{\mu\nu}\!;\lambda}(\delta\Gamma^\lambda_{\mu\nu})
-R^{\mu\nu}_{\phantom{\mu\nu}\!;\nu}(\delta\Gamma^\lambda_{\mu\lambda})
\Bigr\} \sqrt{-g}\,d^4x \;\rightarrow 
\nonumber \\ 
&&\rightarrow \:\;
\alpha_2\!\int \!
\Bigl\{[R^{\mu n}_{\phantom{\mu\nu}\!;\lambda}] 
	g^{\lambda\varkappa}(\delta g_{\varkappa\mu}) +
[R^{n\nu}_{\phantom{\mu\nu}\!;\lambda}]g^{\lambda\varkappa}
(\delta g_{\varkappa\nu})  \nonumber \\
&&+
 \epsilon\,
[R^{\mu \nu}_{\phantom{\mu\nu}\!;n}](\delta g_{\mu\nu}) +
[R^{n\nu}_{\phantom{\mu\nu}\!;\lambda}
]g^{\lambda\varkappa}(\delta g_{\varkappa\nu})
\Bigr\} \sqrt{|\gamma|}\,d^3x. 
\end{eqnarray}
We do not intend to show all the details of very long and cumbersome calculations, and present here only the final result for  $\delta S_{\rm gr}(\alpha_2)$: 
\begin{eqnarray}
\label{deltaSgralpha2}
\delta S_{\rm gr}(\alpha_2)&=&
\alpha_2\!\int\limits_{\Sigma_0} \! \Bigl\{ \!\!
-2(g^{il}g^{jp}+g^{ij}g^{lp})[K_{lp,n} ] 
(\delta K_{ij})  \nonumber \\
&+&
 \!\!\epsilon\,(K^{lp}+Kg^{lp})[K_{lp,n}](\delta g_{nn}) +
2(g^{il}g^{jp}+g^{ij}g^{lp})[K_{lp,n|j} ] (\delta g_{in}) 
\nonumber \\
&+& \Bigl(-(g^{il}g^{jp}+g^{ij}g^{lp})[
K_{lp,nn}]+(-4g^{il}K^{jp}+g^{lp}K^{ij} \nonumber \\
&-&5g^{ij}K^{lp}+(g^{il}g^{jp}+g^{ij}g^{lp})K)[
K_{lp,n}]\Bigr) (\delta\gamma_{ij}) \Bigr\} \sqrt{|\gamma|}\,d^3x. 
\end{eqnarray}

\subsection{$\alpha_3$ --- patch}
\label{alpha3patch}

Calculations of the $\alpha_3$ --- patch are much more simple. As before the first step yields
\begin{eqnarray}
\label{deltaSgralpha3}
\delta S_{\rm gr}(\alpha_3)&=& 2\alpha_3\!\int \!\! 
Rg^{\mu\nu}(\delta R_{\mu\nu}) \sqrt{-g}\,d^4x. 
\nonumber \\
&=&
-2\alpha_3\!\int\limits_{\Sigma_0} \! [R] \Bigl\{g^{\mu\nu}(\delta\Gamma^n_{\mu\nu}) - 
g^{\mu n}(\delta\Gamma^\lambda_{\mu\lambda})
\Bigr\} \sqrt{|\gamma|}\,d^3x. \nonumber \\
&=&
-2\alpha_3\!\int\limits_{(\pm)} \! \Bigl\{
R_{;\lambda}\Bigl(g^{\mu\nu}(\delta\Gamma^\lambda_{\mu\nu})
-g^{\mu\lambda}(\delta\Gamma^\nu_{\mu\nu})\Bigr) 
\Bigr\} \sqrt{-g}\,d^4x.
\end{eqnarray}
The surface integral is transformed into
\begin{eqnarray}
\label{deltaSgralpha3surf}
&&
4\alpha_3\!\int\limits_{\Sigma_0} \! \Bigl\{ -2g^{ij}g^{lp}
[K_{lp,n}] (\delta K_{ij})+\epsilon Kg^{lp}
[K_{lp,n}]  \nonumber \\
&+& g^{ij}g^{lp}[K_{lp,n|j}] (\delta g_{ni})+K^{lj}g^{lp}
[K_{lp,n|j}] (\delta\gamma_{ij})\Bigr\} \sqrt{|\gamma|}\,d^3x.
\end{eqnarray}
The contribution of the remaining volume integrals to the surface integral $\Sigma_0$ equals
\begin{equation}\label{deltaSgralpha3fin}
2\alpha_3\,\epsilon\!\int\limits_{\Sigma_0} \!
\Bigl\{ [R_{;l}] g^{li}(\delta g_{in})-g^{ij}[R_{;n}] 
(\delta\gamma_{ij})\Bigr\} \sqrt{|\gamma|}\,d^3x.
\end{equation}
And the final result is 
\begin{eqnarray}
\label{deltaSgralpha3fin}
\delta S_{\rm gr}(\alpha_3)&=&4\alpha_3\!\int\limits_{\Sigma_0} \!
\Bigl\{-2g^{ij}g^{lp}[K_{lp,n}] (\delta K_{ij}) + \epsilon Kg^{lp} 
[K_{lp,n}] (\delta g_{nn})  
\nonumber \\
&+&2g^{ij}g^{lp}[K_{lp,n|j}]
(\delta g_{in})+\Bigr(\!\!-g^{ij}g^{lp}[K_{lp,nn}]
\nonumber \\
&+&(-5g^{ij}K^{lp}+Kg^{lp}g^{ij}+K^{ij}g^{lp})[K_{lp,n}]
\Bigr)(\delta\gamma_{ij})\Bigr\} \sqrt{|\gamma|}\,d^3x. 
\end{eqnarray}

\subsection{Total}

The complete variation of the gravitational  integral for double layer in Quadratic Gravity is equal to
\begin{eqnarray}
\label{deltaSgralpha3fin}
\delta S_{\rm gr}&=&\int\limits_{\Sigma_0} \!
\Bigl\{ \Bigl\{-2\Bigl((4\alpha_1+\alpha_2)g^{il}g^{jp}+(\alpha_2
+ 4\alpha_3)g^{ij}g^{lp}\Bigr)\Bigr\}[ K_{lp,n}] (\delta K_{ij}) 
\nonumber \\
&+&\epsilon\, \Bigl\{(4\alpha_1+\alpha_2)K^{lp} + (\alpha_2+4\alpha_3)Kg^{lp}
\Bigr\}[K_{lp,n}](\delta g_{nn})  
\nonumber \\
&+&2\, \Bigl\{(4\alpha_1+\alpha_2)g^{il}g^{jp} 
+(\alpha_2+4\alpha_3)g^{ij}g^{lp}\Bigr\}[K_{lp,n|j}](\delta g_{in})  
\nonumber \\
&+&\Bigl\{-\Bigl((4\alpha_1+\alpha_2)g^{il}g^{jp}
+(\alpha_2+4\alpha_3)g^{ij}g^{lp}\Bigr)[K_{lp,nn}]  
\nonumber \\
&-&4(4\alpha_1+\alpha_2)g^{il}K^{jp}[K_{lp,n}] 
\nonumber \\
&+& 
\Bigl((4\alpha_1+\alpha_2)g^{il}g^{jp} 
+ (\alpha_2+4\alpha_3)g^{ij}g^{lp}\Bigr)K[K_{lp,n}]
\nonumber \\
&+&\Bigl((\alpha_2+4\alpha_3)(g^{lp}K^{ij}-5g^{ij}K^{lp})\Bigr)
[K_{lp,n}]\Bigr\}(\delta\gamma_{ij}) \Bigr\} \sqrt{|\gamma|}\,d^3x. 
\end{eqnarray}

\section{Field equations for the double layer}
\label{fieldeqs}

In this section we will analyze the results obtained so far, and derive the equations of motion for the double layers.

We already know that the variation of the total action should be zero on the singular hypersurface ${\Sigma_0}$, i.\,e., ${\delta S_{\rm gr}}{\big|_{\Sigma_0}}+{\delta S_{\rm m}}{\big|_{\Sigma_0}}=0$. From this it follows that 
\begin{equation}\label{deltaSgrsigma0}
{\delta S_{\rm gr}}{\big|_{\Sigma_0}}=
\frac{1}{2}\int\limits_{\Sigma_0} \! 
S^{\mu\nu}(\delta g_{\mu\nu})\sqrt{|\gamma|}\,d^3x.
\end{equation}

First of all, we notice that the constants $\alpha_1$, $\alpha_2$ and $\alpha_3$, given in the Quadratic Gravity Lagrangian ${\cal L}_2$, enter the expression for the variation $\delta S_{\rm gr}$ only in two combinations, namely, proportional to $(4\alpha_1+\alpha_2)$ and $(\alpha_2+4\alpha_3)$. If they are simultaneously zero, then $S^{\mu\nu}=0$, i.\,e., 
\begin{equation}\label{array}
\left\{
\begin{array}{l}	
\alpha_2=-4\alpha_1; \\ \alpha_3=\alpha_2.
\end{array}	 
 \qquad \rightarrow \qquad S^{\mu\nu}=0. \right. 
\end{equation}
But this is exactly the combination of the coefficients in the Gauss--Bonnet term. Thus, when Riemann curvature tensor undergoes a jump at some singular hypersurface ${\Sigma_0}$ (what implies automatically the validity of the Lichnerowicz conditions), then the Gauss--Bonnet term produces neither the double layers nor thin shells. Such a conclusion is by no means trivial, because the Gauss--Bonnet term, being topological, does not effect field equations in the bulk, but contributes to the surface integrals at the boundaries. Moreover, as can be checked, the appearance of the jumps in the Christoffel symbols (and, consequently, the appearance of the Dirac's $\delta$-functions in the curvature) does not lead to the $\delta^2$-terms in the Quadratic Gravity Lagrangian. So, in this case the imposing of the Lichnerowicz conditions is not obligatory. The problem with the Gauss--Bonnet term deserve, therefore, further investigation.

Second, let us consider, what happens, if the jumps in curvatures are zero, i.\,e., when  $[K_{lp,n}]=0$, and no double layer exists at all. We see, that most of the terms in $\delta S_{\rm gr}$ disappears, and we are left with the following relation
\begin{eqnarray}
\label{nojumps}
&-& \! \!\int\limits_{\Sigma_0} \! \Bigl\{ 
(4\alpha_1+\alpha_2)g^{il}g^{jp}+(\alpha_2 
+4\alpha_3)g^{ij}g^{lp}\Bigr\}[K_{lp,nn}]
(\delta\gamma_{ij}) \sqrt{|\gamma|}\,d^3x 
\nonumber \\
&=&\frac{1}{2}\int\limits_{\Sigma_0} \! S^{\mu\nu}(\delta g_{\mu\nu})\sqrt{|\gamma|}\,d^3x.
\end{eqnarray}
It follows, then, that
\begin{equation}\label{array2}
\left\{
\begin{array}{l}	
\!\!\!-\Bigl\{ \! (4\alpha_1\!+\!\alpha_2)g^{il}g^{jp}\!
+\!(\alpha_2\! + \!4\alpha_3)g^{ij}g^{lp}\Bigr\}
[K_{lp,nn}] = \frac{1}{2}S^{ij}; \\ 
\!S^{nn}=0, S^{ni}=0; \\
\![ K_{ij,n}]=0.
\end{array}	 \right.
\end{equation}
These are the analog of the Israel equations for the thin shells in General Relativity \cite{Israel66,Israel67,Cruz67}.

Let us come now to the investigation of the generic case, when there is a jump in the curvature at the singular hypersurface  $\Sigma_0$ and there is no pure the Gauss--Bonnet term, i.\,e., when the double layer really exists. We at once encounter the problem. Namely, in the variation $\delta S_{\rm gr}$ we have the term proportional to the variation of the extrinsic curvature tensor, $\delta K_{ij}$, while in variation $\delta S_{\rm m}$ it is absent by definition.
What to do? The solution is in recognizing that $\delta K_{ij}$ are not the independent variations. Surely, $\delta K_{ij}$ depend on  $\delta\gamma_{ij}$, simply because $\delta K_{ij}=-(1/2)(\delta\gamma_{ij,n}){\big|_{\Sigma_0}}$. But, in a sense, the relation between them is arbitrary, since the equations in the bulk, i.\,e., in $(\pm)$-regions are of the fourth order in derivatives of the metric tensor, and they are not uniquely defined by $g_{\mu\nu}$ and $g_{\mu\nu,\lambda}$ at some Cauchy hypersurface. Thus, we are forced to demand
\begin{equation}\label{Bij}
 \delta K_{i'j'}=B_{i'j'}^{\phantom{ijj}ij}(\delta\gamma_{ij}).
\end{equation}
The appearance of the arbitrary function is not absolutely surprising, though 
it was not expected at the beginning of our investigation. This is a reminiscent of the  $\delta'$-function in the field equations and, thus, it is a marker for the double layer. Indeed, let us consider an equation
\begin{equation}\label{Bij}
A_1(n,x)\delta'(n)+A_2(n,x)\delta(n)+\ldots=A_3(n,x)\delta(n)+\ldots.
\end{equation}
Following the rules of the theory of distributions one should multiply it by an arbitrary function, say $f(n,x)$, with compact support, and then integrate over the variable $n$. The result is
\begin{equation}\label{A}
\!\!\!\!\!\!\!\!-\frac{\partial f}{\partial n}(0,x)A_1(0,x)-f(0,x)\frac{\partial A_1}{\partial n}(0,x)\!+\!f(0,x)A_2(0,x)\!=\!f(0,x)A_3(0,x).
\end{equation}
Dividing then by f(0,x), one gets
\begin{equation}\label{A1}
\varphi(x)A_1(x)-\frac{\partial A_1}{\partial n}(x)+A_2(,x)\!=\!A_3(x), \quad  \varphi(x)=-\frac{1}{f(0,x)}\frac{\partial f}{\partial n}(0,x).
\end{equation}
Only now we are able to write down the equations for the double layers in the Quadratic Gravity.
\begin{eqnarray}
\label{QG1}
&&\epsilon\,\Bigl\{(4\alpha_1+\alpha_2)K^{lp}
+(\alpha_2+4\alpha_3)Kg^{lp}\Bigr\} [K_{lp,n}]
= \frac{1}{2}S^{nn},  \\
\label{QG2}
&&2\,\Bigl\{(4\alpha_1+\alpha_2)g^{il}g^{jp}
+(\alpha_2+4\alpha_3)g^{ij}g^{lp}\Bigr\} K_{lp,n|j}]
= \frac{1}{2}S^{in},  \\ 
&&\Bigl\{\Bigl(-2(4\alpha_1+\alpha_2)g^{i'l}g^{j'p} + (\alpha_2+4\alpha_3)g^{i'j'}g^{lp} \Bigr)[ K_{lp,n}]
B_{i'j'}^{\phantom{ijj}ij} 
\nonumber \\
\label{QG3}
&&+\Bigl\{-\Bigl((4\alpha_1+\alpha_2)g^{il}g^{jp}
+(\alpha_2+4\alpha_3)g^{ij}g^{lp}\Bigr)[K_{lp,nn}]  
\nonumber \\
&&-4(4\alpha_1+\alpha_2)g^{il}K^{jp}[K_{lp,n}] 
\nonumber \\
&&+ 
\Bigl((4\alpha_1+\alpha_2)g^{il}g^{jp} 
+ (\alpha_2+4\alpha_3)g^{ij}g^{lp}\Bigr)K[K_{lp,n}]
\nonumber \\
&&+\Bigl((\alpha_2+4\alpha_3)(g^{lp}K^{ij}-5g^{ij}K^{lp})\Bigr)
[K_{lp,n}]\Bigr\}(\delta\gamma_{ij}) \Bigr\} = \frac{1}{2}S^{ij}.
\end{eqnarray}
Unlike in General Relativity, $S^{nn}$ and $S^{ni}$ are not necessary zero. This fact was first discovered and emphasized by J. M. M. Senovilla.

Here, the singular hypersurface $\Sigma_0$ was considered as given a priori. But, in applications, we are dealing with the situation, when the solution in the $(\pm)$-regions are given, and our task is to find the singular hypersurface $\Sigma_0$, where they may be matched. In the case of the timelike hypersurface it means that we are looking for the trajectory of the double layer. Then, the whole set of of the field equations on $\Sigma_0$ can be divided into two quite different parts. The $(nn)$ and $(ni)$ equations, together with the Lichnerowicz conditions, are needed for determining the surface $\Sigma_0$, while $(ij)$ equations serve for determining the ``arbitrary'' function $B_{i'j'}^{\phantom{ijj}ij}$.

\section{Conclusion and Discussion}

The aim of the present paper was, by using the least action principle only,   to derive the matching conditions for solutions to the field equations of Quadratic Gravity in two bulk region separated by a singular hypersurface.
While in General Relativity the matching surface is called singular, when the matter energy-momentum tensor is concentrated on it, i.\,e., it has a  $\delta$-function term, and as a consequence, the extrinsic curvature tensor has a jump there, resulting in the appearance of the $\delta$-function term in curvatures, the situation in Quadratic Gravity is more subtle. If one assumes the existence of the $\delta$-term in curvature, this would mean the appearance of the $\delta^2$-term in the Lagrangian in Quadratic Gravity (in generic case), what is forbidden in the conventional theory of distributions. Then, the curvature may have only a jump on the matching surface. Since the field equations are now of the second order in derivatives of curvatures (of the fourth order in derivatives of the metric tensor), this means that the left-hand-side (gravitational) of the equation will have terms proportional to $\delta$-function and its derivative, $\delta'$-function. In such a case the singular matching surface is called the double layer.

The $\delta'$-term in the field equations leads to the very interesting phenomenon. After integration in the direction, normal to the singular hypersurface, there appear arbitrary functions in the matching conditions, unlike the General Relativity (all the details are in the preceding Section~\ref{fieldeqs}). But, we were very much surprised when recognized that in our approach the $\delta'$-function is not even mentioned, and the $\delta$-functions themselves are just mentioned, the different terms containing them, being canceled prior to integration. Then, where the arbitrary functions may come from? We solved this puzzle. It appeared that, in the process of variation of the action integral, after implementation of the Stokes' theorem, we are left not only with the variations of the metric tensor on the singular hypersurface, which one needs, but also with the variations of the extrinsic curvature tensor, which one does not need. These two types of variations are not, in fact independent, they are both induced by variations of the solutions in the bulk. And they are connected by some functions, which  are not completely arbitrary, but have some functional freedom that can be removed in the process of solving the whole problem: solutions in the bulk plus matching conditions. Thus, the nature of these ``arbitrary functions'' becomes quite clear.

The structure of the equations for the double layer is the following. There are 6 equations (\ref{QG3}) that form a $3$-dimensional symmetric tensor. The right-hand-side of these equations is the surface energy-momentum tensor of the matter concentrated on our singular hypersurface, i.\,e., the thin shell. In General Relativity these equations serve for the determination of the thin shell trajectory. In Quadratic Gravity their role is different, they serve for the determination of the ``arbitrary'' tensor functions, specific for every choice of the solutions in the bulk regions. The other set of equations consists of scalar $(nn)$ (equation (\ref{QG1})) and $3$-dimensional vector $(ni)$ ones (equation (\ref{QG2})). They serve, given the solution in the bulk, for determining the trajectory of the double layer. The right-hand-side of these equations are $S^{nn}$ and $S^{ni}$ coefficients of the term with $\delta$-function in the energy-momentum tensor. They were discovered by J. M. M. Senovilla, who emphasized their importance and called, correspondingly, ``the external pressure'' and ``external flow''. Evidently, they are not the components of the surface energy-momentum tensor of the thin shell. Nevertheless, their origin is in the matter Lagrangian. We guess that $S^{nn}$ and $S^{ni}$ may appear responsible for the matter field creation by the double layer, and the ``external flow'' will bring the energy out, thus destroying the ``creator''. In General Relativity these entities are zero by virtue of the field equations.

The above speculation is based implicitly on the assumption that the singular hypersurface ($=$ double layer) is time-like. But it could be space-like as well. In General Relativity the space-like hypersurfaces ($=$ thin shells) were used for a phenomenological description of the  cosmological phase transitions \cite{BerKuzTkach1,BKT83b,BerKuzTkach4} and for the abrupt transition to the de Sitter phase inside the black holes \cite{FrolMarMuk,FrolMarMuk2}. In Quadratic Gravity they may appear to be an adequate description of the creation of the Universe from the black hole singularity.

In the present paper we confined ourselves to the time-like and space-like double layers. Surely, there can exist also the null double layers. But there consideration requires quite different mathematical tools and will be done separately.

Two special cases are of particular interest. One of them is the famous Gauss-Bonnet term. It is topological, i.\,e., does not effect the field equations in the bulk and contributes only to the boundary surfaces. Therefore, it can produce its own double layers and thin shells, absent in General Relativity without adding the Gauss-Bonnet term to the Hilbert Lagrangian. Our result above that it is not the case provided the Lichnerowicz conditions are imposed. But, it can be easy checked that, when configuration of the curvatures in Quadratic Gravity  form just the Gauss-Bonnet term, the $\delta^2$ does not appear in the Lagrangian, so the Lichnerowicz conditions are not obligatory. Therefore, such a situation requires further investigation.

The other special case if one has solely the $\alpha_3$-term in the quadratic part of the  Lagrangian. This is the Starobinsky inflationary model and, at the same time, particular case of the more general $F(R)$ theory. Such a theory (originally in the so-called Jordan frame) can be reformulated, by the use of the specific conformal transformation, to the Einstein$+$scalar theory (the so-called Einstein frame). Thus in the Jordan frame one has the Lichnerowicz conditions on the singular hypersurface, only jumps in the curvatures and ``arbitrary'' functions in the equations for the double layer, while in the Einstein frame --- no double layers (only thin shells), no ``arbitrary'' functions and the $\delta$-functions in the curvatures. The seemingly controversial situation is explained very simply. These two incarnations are equivalent only in the bulk, up to the surface terms in the action integral. In the Einstein frame there is an extra degree of freedom, the scalar field as the new dynamical variable, what makes it possible to transform theory with the fourth order derivatives of the metric tensor into that one with only the second derivatives. And, what is crucial for our consideration, that the conformal transformation involved, has the jump at the singular hypersurface. It is this very jump that causes both the appearance of the $\delta$-function in the conformally transformed curvature scalar and the disappearance of the ``arbitrary'' functions, depending on the choice of the solutions in the bulk, replacing them by the jump in the scalar field, depending on the choice of the solution in the bulk.

Some preliminary results were obtained in \cite{bde19,bdes20}.

\ack

This work was supported in part by the Russian Foundation for Basic Research project 18-52-15001-NCNIa.

\end{document}